\documentclass[aps,preprint]{revtex4}
\usepackage{graphicx}
\begin{document}
\title{The origin of power-law distributions in self-organized criticality}
\author{C.B. Yang}
\affiliation{Institute of Particle Physics, Hua-Zhong Normal University,
Wuhan 430079, P.R. China}

\begin{abstract}
 The origin of power-law distributions in self-organized criticality
 is investigated by treating the variation of the number of active sites
 in the system as a stochastic process. An avalanche is then
 regarded as a first-return random walk process in a one-dimensional
 lattice. Power law distributions of the lifetime and spatial size are found
 when the random walk is unbiased with equal probability to move in opposite directions.
 This shows that power-law distributions in self-organized criticality may be caused
 by the balance of competitive interactions. At the mean time, the mean spatial size for
 avalanches with the same lifetime is found to increase in a power law with the lifetime.
\end{abstract}

\maketitle
Power-law distributions are found in a variety of studies from the
populations in cities all over the world, word frequencies
in literature \cite{zipf}, the strength of earthquakes \cite{tremb},
the wealth of individuals \cite{mbr},the forest fire \cite{forest},
distinction of biological species \cite{ecol}, to web site page etc. \cite{more}.
Though the power-law distributions seem to be ubiquitous, the dynamical origin
of them is not very clear up to now. Seventeen years ago, Bak and coworkers studied a
sandpile model \cite{btw} and found power-law distributions for the size and lifetime
of avalanches when the system is in a self-organized critical state. Further studies
\cite{soc1,soc2} on self-organized critical systems show that the power-laws of the spatial
and temporal size are fingerprints of the self-organized criticality (SOC). Although
different authors claimed power-law distributions in different studies, the microscopic
criterion for the onset of SOC and the appearance of power-law distributions
is still lacking.

In this paper we focus our attention on the common features in all SOC models and try to
find out the origin of the power-laws associated with different SOC states. The observation
in this paper is that the dynamics of a SOC system, no matter whether
it is deterministic or random, can be described by the variation of the number
of some kind active sites. Here an active site can be a site with local slope larger
than the threshold in the sandpile model \cite{btw}, or a site with fitness smaller than
the gap in the Bak-Sneppen (BS) model \cite{bs}, etc. In the evolution of the specified system
the number of active sites can increase or decrease according to specific dynamical rules of
the model. For most of known SOC models, the probability for the system to have a large
number of active sites increases with the evolution of the  system. To see this, one can consider
the BS model as an example. In the early updates, because the gap is small, the newly updated
(uniformly distributed) random fitness on the involved sites is more likely to be larger than the
gap. The tendency to decrease the number of active sites dominates.
Therefore, the probability for the system having a
large number of active sites is small, and the mean lifetime of the so-called $f_0$-avalanches
\cite{bak96} is also small for small $f_0$. When
the system's gap becomes larger and larger, the system can increase the number of active
sites with larger and larger probability, and the mean lifetime of $f_0$-avalanches
becomes longer and longer. When the mean lifetime turns out to be infinity,
the system reaches a stationary state--the critical state. Similar statements
can be made for the sandpile model. One can see that, on the variation of the number
of active sites, there are two opposite trends in the interactions in the system for
different SOC models. One trend is
to increase the number of active sites, due to the increase of the gap in the BS model, or
due to adding a grain of sand to the system in the sandpile model or toppling grains to
nearby sites at the threshold. The other is to decrease the number, due to the
larger updated fitness in the BS model or redistributing the grains
to nearby sites far below the threshold in the sandpile model.

From the above discussions one can guess that the power-law distributions associated with
the critical state may result from the balance of the two competitive trends. To confirm
this guess, one can focus on the variation of the number of active sites in the model.
In fact, an avalanche can be defined as a process from a state with at least one active
site to one without active site. Although different models have different rules for the evolutions
or the fluctuation behavior of the active sites, the emergence of power-law distributions
seems to be independent of the dynamical detail.
Here we can consider the simplest trivial ``dynamical'' rules by
treating the variation of the number of active sites as a biased random walk
in a one-dimensional discrete lattice. Similar stochastic description has been used in other SOC
models \cite{mf,lubeck}. Our description is different from the mean-field
approach \cite{mf} in that we do not assume independent evolution for each site.
In fact, we will not assume any specific evolution rule for each site. Rather, we
focus on the evolution of the total number of active sites.
In this paper the distance from the origin is regarded as
the number of active sites of the system. Initially the system has no active sites,
the walker is at the origin. In each update the system may gain one active site
(the random walker moves right) with probability $f_0$ or lose one active site
(the walker moves left) with probability $1-f_0$. If after an update the walker
returns to the origin for the first time, an avalanche is over with the number of
steps defined as its lifetime and the maximum distance from the origin as its spatial size.
Then one can obtain the distributions of the lifetime and the spatial size of avalanches from
analytical calculation and simple computer simulations for each fixed $f_0$. We would like to
investigate whether the random walk mechanism can give rise to the power-law distributions.
Even without any calculation one can see easily from the random-walk based definition
of the avalanche that the probability of having a long lifetime avalanche is extremely
small if $f_0$ is very small and that there is a considerably nonzero probability for an
avalanche having infinite lifetime if $f_0$ is large enough. Therefore, one cannot expect
power-law distribution for the lifetime of avalanches if $f_0$ is either too small or too
large. The possible power-law distribution can be seen only for $f_0$ close to 0.5. It is
our task of this paper to see whether some power-law behaviors can be observed at some
specific $f_0=f_C$ and to find the value of $f_C$ if it is unique.

\begin{figure}
\includegraphics[width=0.45\textwidth]{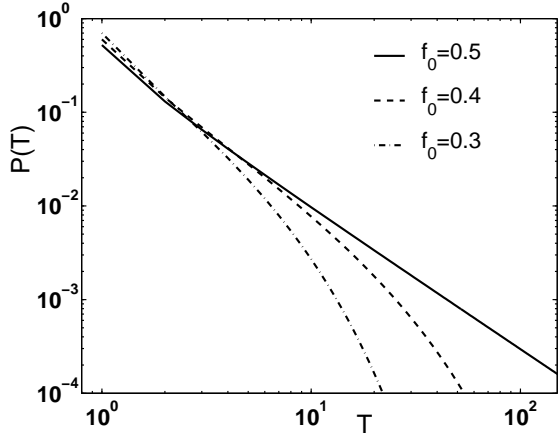}
\caption{Lifetime distribution calculated from Eq. (\ref{eq:theo}). The distribution
is normalized to 1. The slope of the linear fit in log-log plot is $-1.55$ for the case
with $f_0$=0.5, in agreement with analytical result in \cite{boer}.}
\label{theo}
\end{figure}

Due to the setting of our problem, the walker can return to the starting point only after
even steps. Let $Q(T)$ be the probability for the walker to return to the
starting point after $2T$ steps. It can be easily seen that
\begin{equation}
Q(T)={2T\choose T} \left(f_0(1-f_0)\right)^T\ .
\end{equation}
$Q(T)$ above contains contributions from two or more first-return smaller processes.
Denote the probability for an avalanche to have lifetime $2T$ by $P(T)$.
It is obvious that $P(T)$ is equal to the probability for the walker to return to
the starting point {\em for the first time} in $2T$ steps.
By decomposing the whole (big) return process into a first-return process and a
subsequent (smaller) normal return process, we can write $P(T)$ in terms of $Q(T)$ as
\begin{equation}
P(T)=Q(T)-\sum_{j=1}^{T-1} P(j) Q(T-j)\ .
\label{eq:theo}
\end{equation}
From this relation, the first-return probability $P(T)$ can be calculated.
The calculated (after normalization) lifetime distribution $P(T)$ is shown in
Fig.\ref{theo} for $f_0=0.3, 0.4, 0.5$. In the same figure, results from Monte
Carlo simulation of the random walk are also presented with the same $f_0$'s.
Excellent agreement can be seen between results from theoretical
calculation and Monte Carlo simulations, as should. For the cases with $f_0=0.3$ and 0.4
the distributions are more likely to be an exponential form. At $f_0=0.5$ all points
except for $T=1$ lie on a straight line in the double-log plot.
Therefore, a power-law distribution of the lifetime of avalanches
is found for $f_0=f_C=0.5$. For this special case, analytical expressions for both
$P(T)$ and $Q(T)$ exists \cite{boer,book}. For $T\to\infty$ the exponent for $P(T)$
is $-1.5$.

To get the distribution of the number of distinct sites visited by the random walker,
one needs to know the conditional probability $P(S|T)$ for the walker visiting $S$ different
sites in $2T$ steps. The investigation based on this conditional probability is called
the run time statistics by some authors \cite{rts}. It is obvious that $P(1|T)=\delta_{1T}$.
First of all, every path the walker moves along in the
first-return problem has equal probability, since the walker moves forward
$T$ steps and backward also $T$ steps. The probability for the occurrence of each such path
is therefore $(f_0(1-f_0))^T$. Thus, the conditional probability $P(S|T)$ must be proportional
to the number $N(S,T)$ of different paths the walker can move in $2T$ steps when the furthest
visited site is $S$. Secondly, one can see that every path starting from and returning back
to the origin can be decomposed into a large loop, $0\to 1\to 2\to \cdots
\to S-1\to S\to S-1\to \cdots \to 2\to 1\to 0$, and some smaller loops nested on the large loop.
Those small loops must not involve the origin, otherwise the walker has returned to the starting
point. The big loop costs $2S$ steps. If every small loop involves only two sites, loop
$i\to i+1\to i$ for example, there are $T-S$ such two-site loops, and each loop can start from
site $i=1, 2, \cdots, S-1$. There may have some loops involving more than two sites, $i\to i+1\to
i+2\to i+1\to i$ for example. This loop takes four steps. If these four steps are used to have
two smallest loops among the three sites, there are three different partitions of the two two-site
loops. Similarly, a four-site loop can correspond to ten partitions of three two-site loops.
If we neglect the small contributions from partitions with more than one three- (or more)
site loops, the total number of partitions is proportional to that with only two-site loops,
and the proportional factor will be cancelled in the final result Eq. (\ref{corr}) for
$P(S|T)$. So, in the following, we only consider two-site loops.
When counting the number of different partitions of two-site loops, one needs to be careful
to distinguish loops at the same position but at different moments. Since only two-site loops
are considered, the location of later loops has certain constraint. If the loops appear before the
furthest site $S$ is reached, the first loop can start at site $i_1=$1, 2, $\cdots$, $S-1$,
but the second loops can start only at a site $i_2=i_1, i_1+1, \cdots, S-1$. Because of this
constraint, the number of different partition of $r$ loops is
\begin{equation}
M(S,r)={r+S-2\choose r}\ .
\end{equation}
Similarly, one can get the number of partitions of $T-S-r$ after-site-$S$-loops.
Therefore,
\begin{equation}
N(S,T)=\sum_{r=0}^{T-S} M(S, r)M(S, T-S-r)\ .
\end{equation}

For large $S$ and $T$ this is a very large number. So the numerical calculation
of $N(S,T)$ can be done only for small $S$ and $T$'s. Theoretically,
\begin{eqnarray}
P(S|T)=N(S,T)\left/\sum_{S=2}^{T}N(S,T)\right. .
\label{corr}
\end{eqnarray}
With $P(S|T)$ the distribution of sites visited by the walker in the first-return
problem is
\begin{equation}
P(S)=\sum_{T} P(S|T) P(T)\ .
\end{equation}

\begin{figure}
\includegraphics[width=0.45\textwidth]{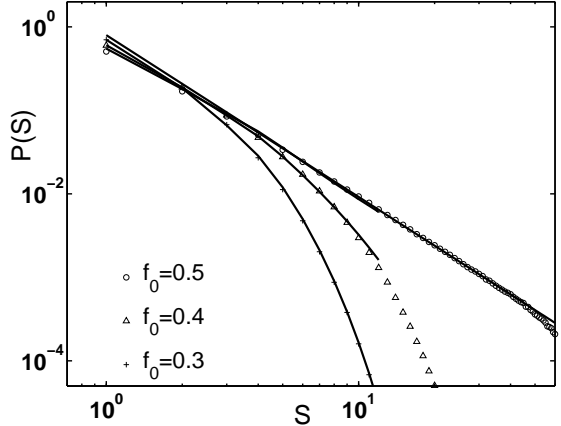}
\caption{Spatial size distribution for first-return random walk problem.
Dots are from Monte Carlo simulation, and the solid curves are from analytical
calculation. The slope of a linear fit in log-log plot for the case with $f_0=0.5$
is $-1.94$.}
\label{theosite}
\end{figure}

The calculated distribution $P(S)$ is shown in Fig. \ref{theosite} for $S$ up to 12. Shown in
the same figure are results from the Monte Carlo simulations for this problem. Theoretical
calculation agree with the Monte Carlo results very well. Once again, power-law
distribution is seen for the case with $f_0=f_C=0.5$ for a wide range of $S$. So, power-law
distributions for the lifetime and spatial size of avalanches can be obtained when the two
competition trends balance each other.

In \cite{mbr} the distribution of the wealth of individuals
is investigated in terms of stochastic process. There the authors claimed that the power-law
distribution can be obtained if one assumes that the relative return of each agent is statistically
equivalent. This equivalence is taken as a sign of market efficiency and is compared to
the Boltzmann distribution in normal statistical systems where some balance is natural and
necessary at equilibrium state. In this paper, we see some balance is also needed for the
appearance of power-law distribution.

It can be mentioned here that the balance of opposite trends is also of necessity
in normal phase transition. Consider the well-known Ising model
for example. There are also two competitive trends in the system.
If there were no thermal fluctuations, all spins would align in the same
direction, because that configuration has the minimum energy.
The thermal motion will destroy the parallel alignment of the spins.
In the low temperature region, the fluctuations are very weak, and
the system is in the ferromagnetic state. At high temperatures, the thermal fluctuations
dominate, and the system is in the paramagnetic state. When the two trends balance,
the system is at the critical point. In the region close a critical point, power-law
behaviors have been observed experimentally.

From above, it is clear that power-law spatial and temporal distributions can be observed when the
two competitive interactions balance. This conforms the conjecture proposed in this
paper and gives the critical value $f_C=0.5$. More interestingly, the exponents obtained
from the first-return random walk problem are quite close to the ones obtained in other
models. This may also indicate that the variation of active sites in different SOC models
can be well approximated by a purely stochastic process.

\begin{figure}
\includegraphics[width=0.45\textwidth]{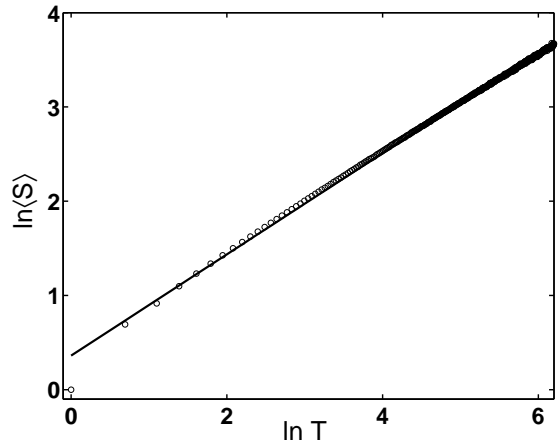}
\caption{Relation between mean spatial size $\langle S\rangle$ and the temporal size $T$
from Monte Carlo simulation. The slope of the linear fit to those points for
$T=2$ to 300 in log-log plot is $0.537$.}
\label{theocorr}
\end{figure}

In fact, one can learn more from such study. One may have noticed that the conditional
probability $P(S|T)$ is universal, independent of $f_0$. From Eq. (\ref{corr})
one can see that there exists correlation between $T$ and $S$. For larger lifetime $2T$
it is more likely for the avalanche having bigger spatial size $S$. In fact, the most probable
spatial size $S$ is about $T/2$ for avalanches of fixed lifetime $2T$.
Quantitatively, one can calculate the mean spatial size for avalanches with
fixed $T$ from Eq. (\ref{corr})
\begin{equation}
\langle S\rangle=\sum_S S\ P(S|T)\ .
\end{equation}
Though we cannot perform analytical or numerical calculation of $\langle S\rangle$
for large $T$, Monte Carlo simulation can be used for this purpose.
The relation between $\langle S\rangle$ and $T$ is shown in Fig. \ref{theocorr}.
A power-law relation can be seen in a very wide range of $T$ from 2 to 100.

It is interesting to note that the average distinct sites visited in $2T$ steps
in conventional random walk is $\langle S\rangle \sim
(16T/\pi)^{0.5}$ for $T\to \infty$. In this paper, only those $2T$ from 2 to 1000
are considered. From Fig. \ref{theocorr} one can see that the points from Monte Carlo
simulation are below the linearly fitted curve for large $T$. In fact, if one does a linear fit
for those points with $T>200$ the slope is less than 0.5. This is understandable,
because in conventional random walk problem the walker may move forward and never return,
thus visit more distinct sites than obtained in this paper.

Finally, we would like to point out that the values of the exponents for the distributions
depend on the dynamical details of the model. But a necessary zero driving rate at the
self-organized critical point implies some balance between the competitive trends.
Therefore, we conclude that in all SOC models power-law distributions result from
the balance of two competitive trends.

As a summary, we studied avalanche dynamics with analytical calculation
and Monte Carlo simulation by treating the variation of the number
of active site in the system as a discrete random process
with probability $f_0$ to increase by 1 and $1-f_0$ to decrease by 1.
An avalanche is defined as a first-return random walk process.
The number of steps of the process is defined as its lifetime of the avalanche, and the number
of different sites the walker has visited is called the spatial size of the avalanche.
Power-law distributions for the temporal and spatial sizes are found $f_0=0.5$. This indicates
that power-law distributions in self-organized critical systems result from the
balance of two competitive trends. The self-organization of the system to the critical state
is a process to adjust the relative strength of the two competitive trends in specific way to
enable them to balance at last. This conclusion may serve as a microscopic criterion for
the onset of self-organized criticality in natural systems.
The correlation between the spatial and temporal sizes of avalanches is also investigated,
and a power-law dependence of the mean spatial size on the temporal size is shown.

This work was supported in part by the National Natural Science Foundation of China
under Grant No. 10075035, by the Ministry of Education of China under Grant
No. 03113. The author would like to thank Prof. R.C. Hwa for his hospitality during his stay
in the University of Oregon where part of this work was done.

\end{document}